\documentclass[twocolumn,showpacs,preprintnumbers,amsmath,amssymb,prb]{revtex4}

\usepackage{graphicx}
\usepackage{dcolumn}
\usepackage{bm}

\begin{document}

\title{Influence of Fermi surface topology \\
on the quasiparticle spectrum in the vortex state}

\author{S.~Graser, T.~Dahm, and N.~Schopohl}

\affiliation{Institut f\"ur Theoretische Physik, Universit\"at T\"ubingen,\\
             Auf der Morgenstelle 14, D-72076 T\"ubingen, Germany}

\date{\today}

\begin{abstract}
We study the influence of Fermi surface topology on the quasiparticle density of
states in the vortex state of type II superconductors. We observe that the
field dependence and the shape of the momentum and spatially averaged density
of states is affected significantly by the topology of the Fermi surface.
We show that this behavior can be understood in terms of characteristic
Fermi surface functions and that an important role is played by the number
of points on the Fermi surface at which the Fermi velocity is directed
parallel to the magnetic field. A critical comparison is made with a
broadened BCS type density of states, that has been used frequently in
analysis of tunneling data. We suggest a new formula as a replacement
for the broadened BCS model for the special case of a cylindrical Fermi 
surface. We apply our results to the two gap
superconductor MgB$_2$ and show that in this particular case the field
dependence of the partial densities of states of the two gaps behaves
very differently due to the different topologies of the corresponding
Fermi surfaces, in qualitative agreement with recent tunneling experiments.
\end{abstract}



\maketitle

\section{Introduction}
\label{intro}

The quasiparticle density of states is one of the characteristic properties
of a superconductor. It contains information about the gap function and
can be probed by various experimental techniques such as tunneling spectroscopy,
photoemission, specific heat, nuclear magnetic resonance (NMR) etc.
These techniques are often used to extract information about the
superconducting gap in a given system. Recently, there is growing interest
to study the quasiparticle excitations in the vortex state of type II
superconductors, particularly in unconventional superconductors.
\cite{Izawa,Hogenboom,MakiTW,Gonnelli,Bugoslavsky,Ekino,Naidyuk} For example,
the zero energy density of states in a $d$-wave superconductor is expected
to vary as the square-root of the magnetic field $B$ instead of the linear
$B$-field variation expected in a dirty $s$-wave 
superconductor.\cite{Volovik,MolerPRL,Revaz} However,
the relation between the gap function and the quasiparticle density of states
is less direct in the vortex state than in the Meissner state. The reason is that
there are important modifications of the density of states due to bound
states in the vortex cores and the
supercurrents running around the vortices, particularly at higher magnetic
fields. In this case quasiparticles are excited with respect to the local
supercurrent flow and their energy appears 'Doppler-shifted'.\cite{Maki} 
In analysis of tunneling spectra these effects are often modeled as a 
'smearing effect' of the zero field density of states.
\cite{Gonnelli,Bugoslavsky,Naidyuk}
However, as we will show below, the
quasiparticle excitations above the moving supercurrent contain information
about the momentum dependent Fermi velocity, especially its direction 
relative to the applied magnetic field. For this reason the particular 
shape of the Fermi surface has an important influence on the momentum
averaged quasiparticle density of states in the vortex state.

\begin{figure}[t]
  \begin{center}
    \includegraphics[width=0.48\columnwidth]{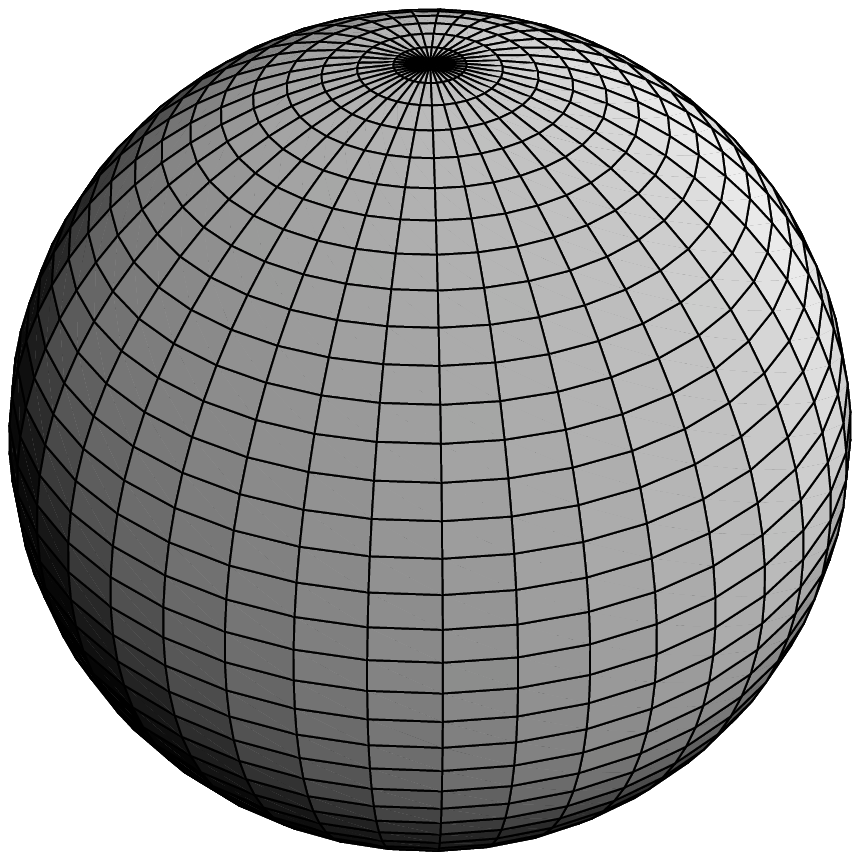}
    \includegraphics[width=0.48\columnwidth]{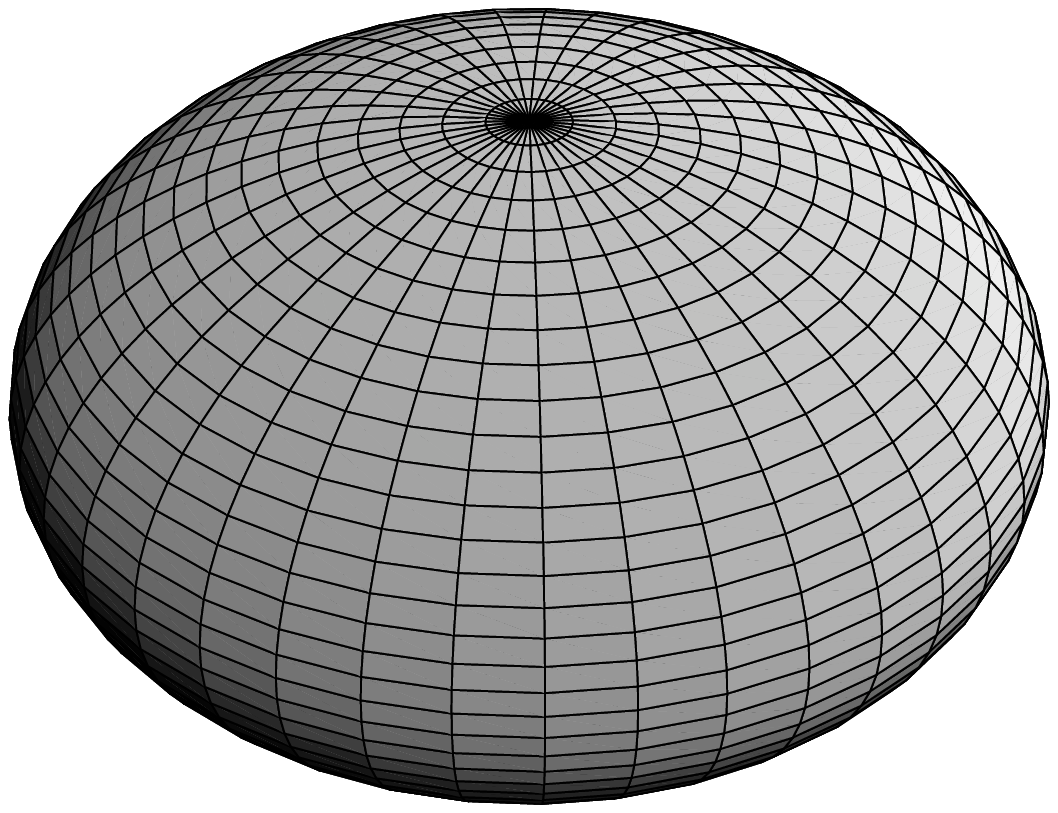}
    
    \vspace{-.4cm}
    \includegraphics[width=0.42\columnwidth]{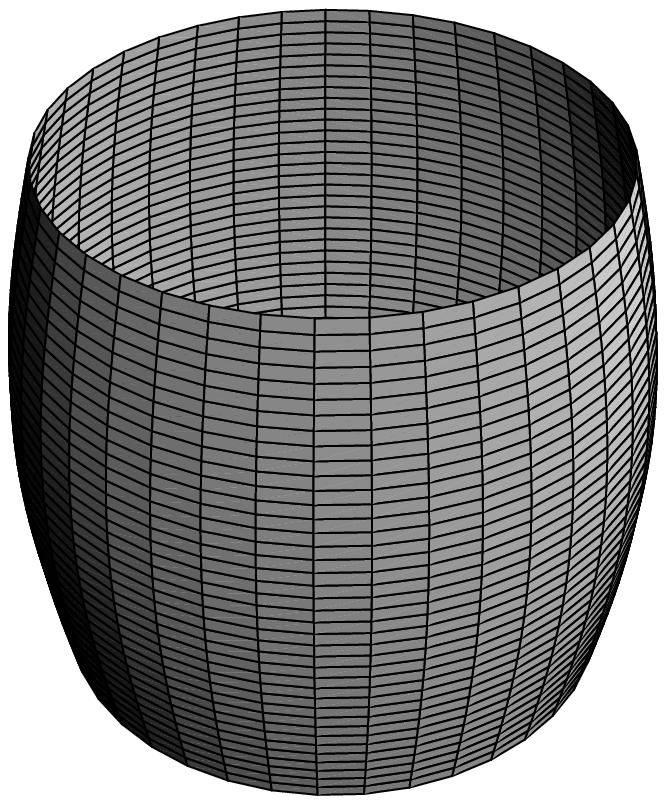}
    \includegraphics[width=0.48\columnwidth]{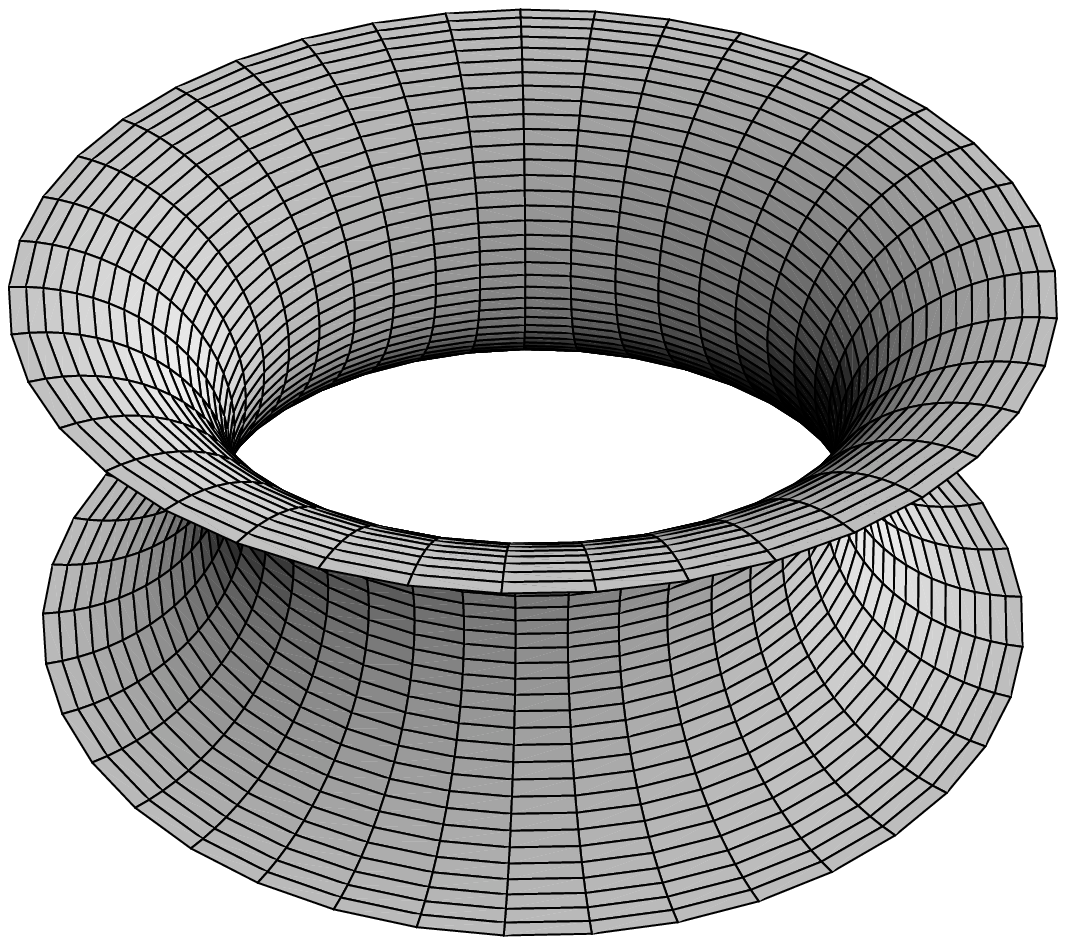}
    \caption{Different shapes of the Fermi surface that are investigated
          within this work. In the upper row we have on the left hand side a spherical 
	  Fermi surface and on the right hand side an elliptical Fermi surface. 
	  In the lower row we show on the left hand side a distorted cylinder  
          with a distortion 
	  parameter $\epsilon_c = 0.163$ and on the right hand side a half-torus. 
     \label{figfermisurf} }
  \end{center}
\end{figure} 

Here, we want to present a detailed study of this influence of the
Fermi surface structure on the quasiparticle density of states in the
vortex state. We are going to compare the different Fermi surface types
shown in Fig.~\ref{figfermisurf}: spherical and elliptical Fermi surfaces
as they are used commonly in isotropic and anisotropic effective mass
models, a cylindrical Fermi surface with a small $c$-axis dispersion
as appropriate for layered
systems, and a Fermi surface of half-torus shape which is relevant for
the $\pi$-band in the recently discussed superconductor MgB$_2$, for
example. \cite{DahmSchopohl1} Our
calculations are based on quasiclassical Eilenberger theory using a method
that was originally introduced by Pesch \cite{Pesch} to describe type II
superconductors at high magnetic fields and that we have generalized to
unconventional pairing symmetry recently.\cite{DGIS} In that work
we have shown that this method leads to much more accurate results
than the frequently used 'Doppler-shift method', particularly at higher
magnetic fields, mainly because it takes into account the contributions from
vortex core states properly. Our results presented below show that the structure of 
the Fermi surface
affects the density of states in the presence of a magnetic field
quite dramatically. We are going to describe our calculational procedure
in section~\ref{spectrum}. In section~\ref{topology} we introduce the
different Fermi surface types used in the present study. 
Section~\ref{integration} contains our results and interpretation in terms
of characteristic Fermi surface functions. In section~\ref{comparison} we
compare our results with the model of a broadened BCS type density of states
that has been used frequently for analysis of experimental data.
In section~\ref{application}
we study the particularly interesting case of the two gap superconductor
MgB$_2$, which possesses four Fermi surfaces of two different types:
two cylindrical $\sigma$-bands and two $\pi$-bands, which can be modeled
by the half-torus in Fig.~\ref{figfermisurf}.\cite{DahmSchopohl1}
These different Fermi surface types lead to very different field
dependencies of the partial densities of states for the two gaps.

\section{\label{spectrum} Quasiparticle spectrum in the vortex state}

Our calculation of the quasiparticle spectrum in the vortex state is based
on quasiclassical Eilenberger theory. \cite{Eilenberger1,LarkinOv1} As
we have shown in Ref.~\onlinecite{DGIS} there exists an accurate approximate
method for the calculation of the density of states averaged over a unit
cell of the vortex lattice (see section IV.A in Ref.~\onlinecite{DGIS}) and
we briefly repeat the essential equations here.
We start from the Eilenberger 
equation for the normal and anomalous component $g$ and $f$ of 
the quasiparticle propagator of a spin-singlet superconductor 
\cite{Eilenberger1,LarkinOv1}

\begin{eqnarray}
\lefteqn{
\left[ 2 \left( i \epsilon_n + \frac{e}{c} \vec{v}_F \cdot \vec{A}
\right) + i \hbar \vec{v}_F \cdot \vec{\nabla} \right] f(\vec{r}, \vec{k}_F, 
i \epsilon_n) = } & & \nonumber \\ 
& & \hspace*{3.3cm} 
2 i g(\vec{r}, \vec{k}_F,i \epsilon_n) \Delta (\vec{r},\vec{k}_F)
\label{eq1}
\end{eqnarray}
Here, $\vec{A}$ is the vector potential and is chosen to be
$\vec{A} = - \frac{1}{2} \vec{r} \times \vec{B}$, the vector $\vec{v}_F = \vec{v}_F
(\vec{k}_F)$ denotes the Fermi velocity at the momentum point $\vec{k}_F$ at
the Fermi surface and $\epsilon_n$ are the fermionic Matsubara frequencies. 
Eq.~(\ref{eq1}) has to be supplemented by a normalization 
condition \cite{Eilenberger1}
\begin{equation}
\left[ g(\vec{r}, \vec{k}_F,i \epsilon_n) \right]^2 + f(\vec{r}, \vec{k}_F, 
i \epsilon_n) f^*(\vec{r}, -\vec{k}_F, i \epsilon_n) = 1
\label{eq2}
\end{equation}
For the spatial variation of the gap function
$\Delta (\vec{r},\vec{k}_F)= \Delta (\vec{k}_F) \cdot \psi_\Lambda (x,y)$
we take the Abrikosov vortex lattice in the following form \cite{DGIS}
\begin{eqnarray}
\psi_\Lambda (x,y) = \frac{1}{\mathcal{N}} \sum_{n=-\infty}^{\infty}
\exp \left[ \frac{\pi \left(ixy - y^2 \right)}{\omega_1 \mbox{Im } \omega_2} 
+ i \pi n + \right. \nonumber \\ 
\hspace{1.3cm} \left. + \frac{i \pi (2n+1)}{\omega_1} (x+ iy) + 
i \pi \frac{\omega_2}{\omega_1} n(n+1) \right]
\end{eqnarray}
where $x$ and $y$ are the coordinates in the plane perpendicular to the
magnetic field $\vec{B}$ and the complex quantities $\omega_1$ and $\omega_2$ 
span the vortex lattice $\Lambda$. $\mathcal{N}$ is a normalization 
factor that has to be chosen in such a way that $|\psi_\Lambda |^2$ averaged over a
unit cell of $\Lambda$ equals unity. The normalized local density of states can be 
obtained from the real part of the analytical continuation of $g$, averaged
over the Fermi surface 
\begin{equation}
N(\vec{r}, E) = \left\langle \mbox{Re} \left\{
g(\vec{r}, \vec{k}_F,E + i 0^+) \right\} \right\rangle_{FS}
\label{eq2b}
\end{equation}
where
\begin{equation}
\Big\langle \cdots \Big\rangle_{FS} = \frac{1}{N(0)} \int_{FS} \frac{d^2 k_F}{(2 \pi)^3}
\frac{1}{|\hbar \vec{v}_F (\vec{k}_F)|} \cdots
\label{FSav}
\end{equation}
denotes an average over the Fermi surface.

Proceeding along the lines in Ref.~\onlinecite{DGIS} we can now find the
following approximate analytical solution for the density 
of states spatially averaged over a unit cell $C_\Lambda$ of the vortex lattice
for arbitrary field directions:
\begin{eqnarray}
\lefteqn{g (\vec{k}_F, i \epsilon_n) =} \label{gkf} \\
& & \frac{1}{|C_\Lambda|} \int_{C_\Lambda} d^2r \;
g(\vec{r},\vec{k}_F, i \epsilon_n) 
= \frac{1}{\sqrt{1 + P_\Lambda(\vec{k}_F, i \epsilon_n)}}
\nonumber
\end{eqnarray}
where the momentum and frequency dependent function 
$P_\Lambda$ takes the form
\begin{equation}
P_\Lambda(\vec{k}_F, i \epsilon_n) = \frac{4 |\Delta (\vec{k}_F)|^2}{|\eta_{k_F}|^2} 
(1-\sqrt{\pi} z w(i z))
\label{Plam}
\end{equation}
The $w$-function, also known as Dawson's integral, is related to the 
complement of the Error function by
\begin{equation}
w(iz) = \frac{1}{i \pi} \int_{-\infty}^{\infty} \frac{e^{-t^2}}{t - iz} dt 
= e^{z^2} \mbox{erfc} (z)
\end{equation}
Furthermore $z$ are normalized Matsubara frequencies of the following form
\begin{equation}
z = \frac{\sqrt{2} \epsilon_n}{|\eta_{k_F}|}
\label{zeps}
\end{equation}
where $\eta_{k_F}$ is proportional to the projection of the Fermi velocity 
into the complex plane perpendicular to the magnetic field direction and
given by
\begin{equation}
\eta_{k_F} = \hbar (v_{F,1} + i v_{F,2}) \sqrt{\frac{e B}{\hbar c}}
\label{etak}
\end{equation}
Here, $v_{F,1}$ and $v_{F,2}$ are the components of the Fermi velocity in the
plane perpendicular to the average magnetic field $\vec{B}$.
Using the flux quantization condition
\begin{equation}
\frac{\pi}{\omega_1 \mbox{Im } \omega_2} = \frac{e B}{\hbar c}
\end{equation}
where $\omega_1 \mbox{Im } \omega_2=|C_\Lambda|$ is the size of a unit cell of the 
vortex lattice (if we assume $\omega_1$ to be real) we can bring $\eta_{k_F}$ into 
the form
\begin{equation}
\eta_{k_F} = \hbar (v_{F,1} + i v_{F,2}) \sqrt{\frac{\pi}{\omega_1 \mbox{Im } \omega_2}}
\end{equation}

If the magnetic field is not directed parallel to the $c$-axis of the uniaxial systems
considered here, then the rotational symmetry around the magnetic field direction is 
broken and we  have to take into account, that the isotropic vortex lattice no longer 
has to be the appropriate ground state. If we use a variational ansatz for a distorted 
Abrikosov vortex lattice with 
\begin{equation}
\psi_\Lambda^\tau (x,y) = \psi_\Lambda (e^{-\tau} x, e^\tau y)
\end{equation}
we find that the Fermi velocity components in Eq.~(\ref{etak}) have to be
replaced by the scaled Fermi velocities \cite{DahmSchopohl1}
\begin{equation}
\tilde{v}_{F,1} = e^\tau v_{F,1}, \hspace{1em} \mbox{and}
\hspace{1em} \tilde{v}_{F,2} = e^{-\tau} v_{F,2}
\label{vfscaled}
\end{equation}
The distortion parameter $\tau$ has to be found by a minimization of the
free energy at a given field strength and temperature.

Within our approximation the free energy difference between superconducting and
normal state is given by the expression
\begin{eqnarray}
\lefteqn{
\Omega_S - \Omega_N = - \Bigg\langle \pi T 
\sum_{|\epsilon_n| < \omega_c}  
\frac{|\Delta(\vec{k}_F)|^2}{\epsilon_n} \sqrt{\pi} z w (i z) }
\label{freeen}  \\ & &
\mbox{Re} \left[ 
\frac{P_\Lambda (\vec{k}_F, i \epsilon_n) }
{\sqrt{1 + P_\Lambda (\vec{k}_F, i \epsilon_n)}
\left( 1 + \sqrt{1 + P_\Lambda (\vec{k}_F, i \epsilon_n)} \right)^2}
\right] \Bigg\rangle_{FS} 
\nonumber
\end{eqnarray} 
A derivation of this expression can be found in Appendix \ref{app1}.
For each value of $\tau$ the gap function in Eq.~(\ref{freeen}) has to be calculated
from the gap equation
\begin{eqnarray}
\left\langle \pi T \sum_{|\epsilon_n| < \omega_c} \frac{|\Delta(\vec{k}_F)|^2}{\epsilon_n} 
\left[ g(\vec{k}_F, i \epsilon_n) \sqrt{\pi} z w (iz) - 1 \right] \right. \nonumber \\
\left. - |\Delta(\vec{k}_F)|^2 \ln \frac{T}{T_c}
\right\rangle_{FS} = 0
\label{gapequ}
\end{eqnarray}
self-consistently. We note that the variational parameter $\tau$ 
enters Eqs.~(\ref{freeen}) and (\ref{gapequ}) only via the
quantity $|\eta_{k_F}(\tau)|$. A determination of $\tau$ requires to
minimize Eq.~(\ref{freeen}), where for each trial value of $\tau$
Eq.~(\ref{gapequ}) has to be solved self-consistently. Fortunately,
we were able to find analytical results for $\tau$ in some of the
cases studied below, such that this minimization procedure could be
avoided in these cases.

The field dependent total density of states averaged over the Fermi 
surface and a unit cell of the vortex lattice is found from Eq.~(\ref{gkf})
\begin{equation}
N(B,E) = \left\langle \mbox{Re} \left\{
\frac{1}{\sqrt{1 + P_\Lambda(\vec{k}_F, E+ i 0^+ )}} \right\} \right\rangle_{FS}
\label{dos}
\end{equation}
We note that the direction of the Fermi velocity enters this equation only 
via the quantity $|\eta_{k_F}|$ and that only the components perpendicular 
to the magnetic field play a role. This is where the Fermi surface structure
comes into play.

\section{\label{topology} Parametrization of the Fermi surface}

In order to parametrize the different Fermi surfaces we have to find an 
appropriate parametrization  of the Fermi wave vector $\vec{k}_F$: 
\begin{equation}
\vec{k}_F (\vartheta, \varphi) = k_x (\vartheta, \varphi) \vec{e}_x
+ k_y (\vartheta, \varphi) \vec{e}_y + k_z (\vartheta, \varphi) \vec{e}_z
\end{equation}
Using this parametrization we can calculate the outward oriented 
normal unit vector $\vec{n}_F$
\begin{equation}
\vec{n}_F = \frac{\vec{t}_\varphi \times \vec{t}_\vartheta}
{|\vec{t}_\varphi \times \vec{t}_\vartheta|}
\end{equation} 
where $\vec{t}_\varphi$ and $\vec{t}_\vartheta$ are two orthogonal tangential vectors
that are parallel to the coordinate lines of $\varphi$ and $\vartheta$
\begin{equation}
\vec{t}_\varphi = \partial_\varphi \vec{k}_F (\vartheta, \varphi), \;
\vec{t}_\vartheta = \partial_\vartheta \vec{k}_F (\vartheta, \varphi)
\end{equation}
The Fermi velocity vector is always directed perpendicular to the Fermi
surface and thus can be written in general as
\begin{equation}
\vec{v}_F = v_F \cdot \vec{n}_F
\end{equation}

The Fermi surface of layered systems as for example the high $T_c$-cuprates
can be described as a distorted cylinder. The distortion is due to a small $c$-axis
dispersion. In this case we can parametrize the distorted cylindrical Fermi surface as
\begin{eqnarray}
k_x (k_c,\varphi) & = &  \left( k_{ab} + \frac{\epsilon_c}{c} \cos (c k_c) \right) 
\cos \varphi, \nonumber \\ 
k_y (k_c,\varphi) & = & \left( k_{ab} + \frac{\epsilon_c}{c} \cos (c k_c) \right) 
\sin \varphi, \nonumber \\
k_z (k_c,\varphi) & = & k_c
\end{eqnarray}
with a dimensionless $c$-axis dispersion parameter $\epsilon_c$. 
The parameter $k_c$ is running from  $-\pi/c$ to $\pi/c$  where $c$ is the 
lattice constant in $c$-axis direction. The polar angle $\varphi$ varies in the 
interval $[0, 2\pi]$. 

Another Fermi surface that we will discuss is a half-torus. In magnesium diboride with its 
hexagonal crystal structure the $\pi$-band possesses a tubular Fermi surface structure that
can be approximated by a half-torus living at the border of the hexagonal unit cell of the 
reciprocal lattice. \cite{DahmSchopohl1}
In this case we have the following parametrization
\begin{eqnarray}
k_x (\vartheta, \varphi) & = &  k_F \cdot (\nu + \cos \vartheta) \cos \varphi, \nonumber \\ 
k_y (\vartheta, \varphi) & = & k_F \cdot (\nu + \cos \vartheta) \sin \varphi, \nonumber \\
k_z (\vartheta, \varphi) & = & k_F \cdot \sin \vartheta
\end{eqnarray}
where the parameter $\nu$ denotes the ratio between the radius of the 
meridian and the radius of the longitudinal curve of the torus. 
Again $\varphi$ can take values between $0$ and $2\pi$ while $\vartheta$ varies
between $\pi/2$ and $3\pi/2$ to parametrize the concave half-torus.
The spherical Fermi surface can be viewed as a 
special case of the toroidal Fermi surface with parameter $\nu = 0$.

If we assume a crystal with different effective masses in the $c$-axis direction and 
within the $ab$-plane 
\begin{equation}
\epsilon_k = \frac{1}{2m_{ab}} (k_a^2 + k_b^2) + \frac{1}{2m_c} k_c^2 - \epsilon_F
\end{equation}
we find an elliptic Fermi surface with the following parametrization
\begin{eqnarray}
k_x (\vartheta, \varphi) & = &  k_F \cdot \cos \vartheta \cos \varphi, \nonumber \\ 
k_y (\vartheta, \varphi) & = & k_F \cdot  \cos \vartheta \sin \varphi, \nonumber \\
k_z (\vartheta, \varphi) & = & k_F \cdot \sqrt{m_c/m_{ab}} \sin \vartheta
\end{eqnarray}
Here, $\vartheta$ takes values between $-\pi/2$ and $\pi/2$. In Fig.~\ref{figfermisurf} 
we show the four different Fermi surfaces discussed above. From these parametrizations
we can now calculate the Fermi velocity and its projection into the 
plane perpendicular to the magnetic field. We will consider the two cases
of a magnetic field applied parallel to the $c$-axis direction of the crystal
and the case of an in plane magnetic field. 

For the cylinder with $c$-axis dispersion $\epsilon_c$ we assume that  
$\epsilon_c^2 \ll 1$. In this case the variation of the in-plane components of the 
Fermi velocity due to the $c$-axis dispersion can be neglected. Then we can write 
\begin{equation}
\vec{v}_F = v_F (\cos \varphi \; \vec{e}_x + \sin \varphi \; \vec{e}_y
+ \epsilon_c \sin c k_c \; \vec{e}_z )
\end{equation}
For the half-torus and the sphere the Fermi velocity is found to be:
\begin{equation}
\vec{v}_F = v_F (\cos \varphi \cos \vartheta \; \vec{e}_x 
+ \sin \varphi \cos \vartheta \; \vec{e}_y + \sin \vartheta \; \vec{e}_z)
\end{equation}
Writing $v_{F,ab} = k_F/m_{ab}$ and $v_{F,c} = k_F / \sqrt{m_c m_{ab}}$ we find 
for the elliptical Fermi surface:
\begin{eqnarray}
\lefteqn{\vec{v}_F = \vec{\nabla}_k \epsilon_k = } \\ 
& & v_{F,ab} ( \cos \varphi \cos \vartheta \: \vec{e}_x 
+ \sin \varphi \cos \vartheta \: \vec{e}_y ) +
v_{F,c} \sin \vartheta \: \vec{e}_z \nonumber
\end{eqnarray}

The Fermi surface average Eq.~(\ref{FSav}) reads in these coordinates
\begin{equation}
\Big\langle \cdots \Big\rangle_{FS} = \frac{c}{4 \pi^2} \int_0^{2\pi} d\phi 
\int_{-\pi/c}^{\pi/c} dk_c \cdots
\end{equation}
for the cylindrical Fermi surface with $\epsilon_c^2 \ll 1$ and
\begin{equation}
\Big\langle \cdots \Big\rangle_{FS} = \frac{1}{2 \pi} \int_0^{2\pi} d\phi 
\int_{\pi/2}^{3\pi/2} d\vartheta \frac{\nu+ \cos \vartheta}{\pi \nu-2} \cdots
\label{FSavtor}
\end{equation}
for the half-torus. The integrations for spherical and elliptical
Fermi surface are obtained from Eq.~(\ref{FSavtor}) for $\nu=0$.

\subsection{\label{caxis} Field in $c$-axis direction}

As pointed out above, for the calculation of the density of states 
in the vortex state using Eq.~(\ref{dos}) we only need to know
the modulus of the quantity $\eta_{k_F}$ introduced in section~\ref{spectrum}.
This quantity only contains the components of the Fermi velocity
perpendicular to the external magnetic field. For field directed
in $c$-axis direction we therefore only need the $ab$-plane components.
As a result of the rotational symmetry of all the Fermi surfaces considered 
we expect no angular dependence on the polar angle $\varphi$. For the 
cylindrical Fermi surface we find no angular dependence at all
\begin{equation}
|\eta_{k_F}| = \hbar v_F \sqrt{ \frac{e B}{\hbar c}}
\end{equation}
For the toroidal or the spherical Fermi surface a dependence on the 
azimuthal angle $\vartheta$ is found:
\begin{equation}
|\eta_{k_F}| = \hbar v_F \left| \cos \vartheta \right| \sqrt{ \frac{e B}{\hbar c}}
\end{equation}
For the elliptic Fermi surface we have
\begin{equation}
|\eta_{k_F}| = \hbar v_{F,ab} \left| \cos \vartheta \right| \sqrt{ \frac{e B}{\hbar c}}
\end{equation}

\subsection{\label{abplane} Field in $ab$-plane direction}

If we consider a magnetic field that is not parallel to the $c$-axis direction, 
the rotational symmetry around the $c$-axis is broken and we find more complicated 
expressions for the Fermi velocity projection $\eta_{k_F}$. In this case
we can find the projection by a simple rotation, for example around the $a$-axis. 
If we carry out a rotation with angle $\gamma$ we obtain
a modified expression for the Fermi velocity in the rotated coordinate frame
\begin{eqnarray}
\vec{v}_F' = v_{F,a} \vec{e}_1 + (\cos \gamma \; v_{F,b} - \sin \gamma \; v_{F,c}) 
\vec{e}_2 \nonumber \\
\hspace{1.5cm} + (\sin \gamma \; v_{F,b} + \cos \gamma \; v_{F,c}) \vec{e}_3
\end{eqnarray}

Here the unit vector $\vec{e}_3$ is chosen to point into the direction of the 
magnetic field and $\vec{e}_1$, $\vec{e}_2$ span the plane perpendicular to the 
magnetic field. $v_{F,a}$, $v_{F,b}$ and $v_{F,c}$ are the components of the Fermi 
velocity along the $a$-, $b$- and $c$-axis of the crystal, respectively. 
If the magnetic field is applied along the $ab$-plane direction of the crystal, the angle 
$\gamma$ equals $\pi / 2$ and the Fermi velocity $\vec{v}_F'$ in the rotated coordinate
frame can be written as
\begin{equation}
\vec{v}_F' = v_{F,a} \vec{e}_1 + (- v_{F,c}) \vec{e}_2 
+ v_{F,b}  \vec{e}_3
\end{equation}
As we have pointed out above, in this field direction the distortion of the
vortex lattice has to be taken into account which leads to a scaling
of the components of the Fermi velocity (see Eq.~(\ref{vfscaled})).
In this way we find for the cylindrical Fermi surface
\begin{equation}
|\eta_{k_F}| = \hbar v_F \sqrt{ \frac{e B}{\hbar c}} \sqrt{e^{2\tau} \cos^2 \varphi 
+ e^{-2\tau} \epsilon_c^2 \sin^2 c k_c} ,
\label{etaabcyl}
\end{equation}
for half-torus and sphere we have
\begin{equation}
|\eta_{k_F}| = \hbar v_F \sqrt{ \frac{e B}{\hbar c}} \sqrt{e^{2\tau} \cos^2 \varphi 
\cos^2 \vartheta + e^{-2\tau} \sin^2 \vartheta} ,
\label{etaabsph}
\end{equation}
and for the elliptical Fermi surface we find
\begin{equation}
|\eta_{k_F}| = \hbar \sqrt{ \frac{e B}{\hbar c}} \sqrt{e^{2\tau} v_{F,ab}^2 \cos^2 \varphi 
\cos^2 \vartheta + e^{-2\tau} v_{F,c}^2 \sin^2 \vartheta}
\end{equation}
In this case it is useful to introduce a rescaling of the distortion parameter 
$\tau$ via
\begin{equation}
e^{\bar{\tau}} = e^{\tau} \sqrt{\frac{v_{F,ab}}{v_{F,c}}}
\end{equation}
Then we can write
\begin{equation}
|\eta_{k_F}| = \hbar \sqrt{v_{F,ab} v_{F,c}} \sqrt{ \frac{e B}{\hbar c}} 
\sqrt{e^{2\bar{\tau}} \cos^2 \varphi 
\cos^2 \vartheta + e^{-2\bar{\tau}} \sin^2 \vartheta}
\label{etaabsell}
\end{equation}
This equation has the same form as Eq.~(\ref{etaabsph}) with $\tau$ being
replaced by $\bar{\tau}$ and $v_F$ by $\sqrt{v_{F,ab} v_{F,c}}$.
In the next section these expressions are used for evaluation of the 
Fermi surface integration and calculation of the spatially and momentum averaged 
density of states.

\section{\label{integration} Fermi surface integration}

In order to calculate the quasiparticle spectrum, we have to integrate the spatially 
averaged density of states over the Fermi surface using Eq.~(\ref{dos}). In the following we
are going to discuss the case of an isotropic $s$-wave gap and the two gap
superconductor MgB$_2$. Some results for a $d$-wave gap and a cylindrical Fermi surface
can be found in our previous work. \cite{DGIS} We note
that for an isotropic $s$-wave gap the function $P_\Lambda$ depends on
momentum $\vec{k}_F$ only via the quantity $|\eta_{k_F}|$. In this case it is
useful to introduce a characteristic Fermi surface function $g_F(s)$ of the following form
\begin{equation}
g_F(s) = \frac{1}{N(0)} \int_{FS} \frac{d^2 k_F}{(2 \pi)^3} 
\frac{1}{|\hbar \vec{v}_F (\vec{k}_F)|} \delta \left( s -  \frac{|\eta_{k_F}|}{\alpha} \right)
\label{gfs}
\end{equation}
For convenience, we have introduced a parameter $\alpha$ containing the field dependence
\begin{equation}
\alpha = \hbar v_F \sqrt{ \frac{e B}{\hbar c}}
\label{alpha}
\end{equation}
For a given value of $s$, $g_F(s)$ counts the number of states at the Fermi surface
with a certain projection of the Fermi velocity into the plane perpendicular to the
magnetic field.
Using the function $g_F(s)$ the density of states for a given value of $\alpha$
can be written as a single integration over the variable $s$:
\begin{equation}
N(B,E) = \int_0^\infty ds \; g_F(s) \;
\mbox{Re} \left[  \frac{1}{\sqrt{1 + P_\Lambda (\alpha s, E)}} \right]
\label{dosgfs}
\end{equation}
with
\begin{equation}
P_\Lambda(\alpha s, E) = \frac{4 |\Delta|^2}{(\alpha \; s)^2} 
(1-\sqrt{\pi} z(\alpha s) w(i z(\alpha s)))
\end{equation} 
Here, $z(\alpha s)$ is the normalized real axis frequency
\begin{equation}
z(\alpha s) = \frac{- i \sqrt{2} }{\alpha  \; s} \left( E + i 0^+ \right)
\end{equation}
All information about the Fermi surface structure is contained in the function
$g_F(s)$, which still depends on the direction of the magnetic field, but not
on the magnitude. Looking at Eq.~(\ref{dosgfs}) $g_F(s)$ can be interpreted as
a weighting function. 

\begin{figure}
  \begin{center}
    \includegraphics[width=0.9\columnwidth]{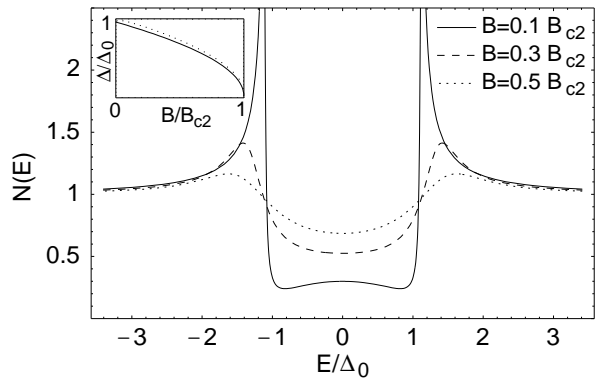}
    \vspace{.4cm}

    \caption{Averaged density of states in the vortex state for the cylindrical 
     Fermi surface with magnetic field in $c$-axis direction for three different
     magnetic fields $B/B_{c2}=$ 0.1, 0.3, and 0.5.
     The inset shows the field dependence of the gap amplitude in units of $\Delta_0$
     obtained numerically from Eq.~(\ref{gapequ}) (dots) and the
     approximation Eq.~(\ref{deltab}) (solid line). 
     \label{ds_cyl} }
  \end{center}
\end{figure} 

\subsection{Field in $c$-axis direction}

The characteristic function $g_F(s)$ assumes a particularly simple form when the
magnetic field is directed along the crystal $c$-axis.
For the cylindrical Fermi surface $|\eta_{k_F}|$ becomes angular independent
and the characteristic function reduces to a $\delta$-function:
\begin{equation}
g_F(s) = \delta \left( s - 1 \right)
\end{equation}
In this case the density of states is just given by 
\begin{equation}
N_{\rm cyl} \left( B,E \right) = 
\mbox{Re} \left[  \frac{1}{\sqrt{1 + P_\Lambda (\alpha, E)}} \right]
\label{doscyl}
\end{equation}
in agreement with Eq.~(36) in Ref.~\onlinecite{DGIS}. This result is shown in
Fig.~\ref{ds_cyl} as a function of the quasiparticle energy 
$E$ in units of the zero field gap value
$\Delta_0=\Delta(B=0) $ for three different field strengths. 
For simplicity we have used a field dependence of the gap function \cite{Maki}
\begin{equation} 
\Delta(B) = \Delta_0 \sqrt{1-\frac{B}{B_{c2}}}
\label{deltab}
\end{equation}
shown as the solid line in the inset of Fig.~\ref{ds_cyl}. This field dependence
is a very good approximation to the field dependence obtained from the gap
equation Eq.~(\ref{gapequ}) numerically (solid squares in the inset).\cite{DahmSchopohl2}
The value of $|\eta_{k_F}|$ at the upper critical field is found from the
linearized gap equation Eq.~(\ref{gapequ}) to be
\begin{equation} 
|\eta_{k_F}|(B_{c2}) = \sqrt{2 \gamma} \Delta_0 \simeq 1.887 \Delta_0
\end{equation}
with $\ln \gamma=0.577215$ being Euler's constant.

\begin{figure}
  \begin{center}
    \includegraphics[width=0.9\columnwidth]{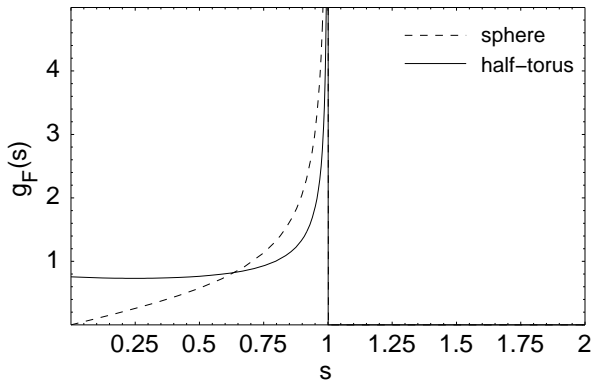}
    \vspace{.4cm}

    \caption{The characteristic functions $g_F(s)$ for the spherical and 
    half-toroidal Fermi surface. The magnetic field is applied along the $c$-axis
    direction and we have chosen the ratio of the toroidal radii to be $\nu = 4$. 
    \label{gFs_c}}
  \end{center}
\end{figure} 

As becomes apparent from Fig.~\ref{ds_cyl},
the peak-to-peak distance of the gap structure in the density of states {\it increases}
with increasing field, while the gap itself {\it decreases}. For this reason the
gap structure seen in the density of states is not a good quantitative measure of
the gap in the vortex state anymore.

Apparently, the density of states in the general case Eq.~(\ref{dosgfs}) can be 
reduced to an integral over the density of states for the cylindrical Fermi surface
Eq.~(\ref{doscyl}):
\begin{equation}
N(B,E) = \int_0^\infty ds \; g_F(s) \;
N_{\rm cyl} \left( \alpha s,E \right)
\end{equation}
This means that $N(B,E)$ consists of a weighted average of
the density of states for the cylindrical Fermi surface in Fig.~\ref{ds_cyl}
for different
magnetic field strengths ranging from $B=0$ (corresponding to $s=0$) up
to a certain upper limit given by the maximum of $|\eta_{k_F}|$ on the
Fermi surface.

For the toroidal Fermi surface the integral over the Fermi surface in Eq.~(\ref{gfs}) 
can be done analytically and for the characteristic function we find
\begin{equation}
g_F(s) = \left\{ \begin{array}{cc} \displaystyle
\frac{\nu - s}{\pi \nu - 2} \frac{2}{\sqrt{1 - s^2}} & {\rm for} \quad 0 \le s < 1 \\[3ex]
0 & {\rm else}
\end{array} \right.
\end{equation}
The result for the spherical Fermi surface is obtained for $\nu = 0$. The same
result is also found for the elliptical Fermi surface, if $v_F$ in
Eq.~(\ref{alpha}) is replaced by $v_{F,ab}$.
The weight functions for the half-torus and the spherical Fermi surface are shown 
in Fig.~\ref{gFs_c}. For the half-torus $\nu=4$ (appropriate for MgB$_2$) has been chosen.
$g_F(s)$ for the half-torus starts with a finite value at $s=0$. This has a topological
reason: on the half-torus with field in $c$-axis direction there exist two lines of 
points at which the Fermi velocity is parallel to the magnetic field. At these
points the projection of the Fermi velocity onto the plane perpendicular to the
field vanishes. In contrast, for the spherical and elliptical Fermi surfaces
there exist only two such points, the poles. Therefore $g_F(s)$ becomes 0 at $s=0$
and increases linearly. For the cylindrical Fermi surface no such points
exist and $g_F(s)$ becomes 0 for $s \neq 1$.

\begin{figure}
  \begin{center}
    \includegraphics[width=0.9\columnwidth]{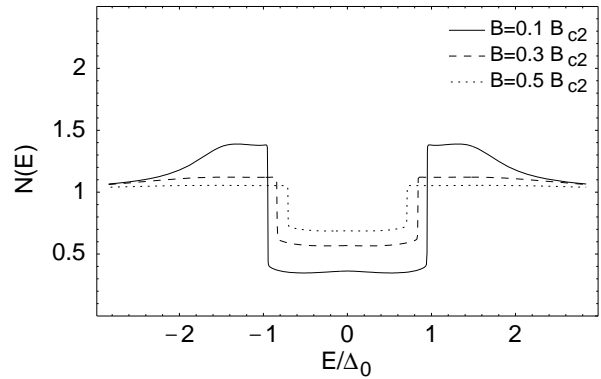}
    \vspace{.4cm}

    \caption{Averaged density of states for the toroidal Fermi surface 
    with magnetic field parallel to the $c$-axis direction
     \label{ds_torus} }
  \end{center}
\end{figure}

In Fig.~\ref{ds_torus} we show the density of states for the half-torus for different
field strengths in $c$-axis direction. Because of the finite value of $g_F(s)$ for
$s=0$ in this case there is large weight to the BCS singularity of the spectrum at
$B=0$. This results in the sharp flank at the
gap edge seen in Fig.~\ref{ds_torus} even for higher magnetic fields. The position of
the sharp flank closely follows the field dependence of the gap. The singularity 
itself is washed out by the averaging process. But still the most important 
contribution comes from the highest field, corresponding to $s=1$ in Fig.~\ref{gFs_c}.

\begin{figure}
  \begin{center}
    \includegraphics[width=0.9\columnwidth]{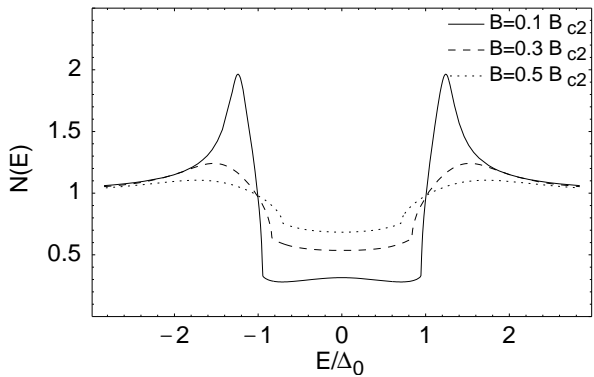}
    \vspace{.4cm}

    \caption{Averaged density of states for spherical and elliptical
     Fermi surface, respectively, \label{ds_sphere} with magnetic field parallel to the $c$-axis direction}
  \end{center}
\end{figure}

The characteristic function of the spherical Fermi surface increases linearly 
from $s=0$. This results in a visible break at the gap edge in the density of states
but no longer in a significant flank, as shown in Fig.~\ref{ds_sphere}. Again, the most 
important contribution comes from the highest fields ($s=1$), leading to a peak-to-peak
distance that increases with increasing magnetic field. The reduction of the
BCS singularity is less pronounced than for the half-torus, because the weight near 
$s=1$ in $g_F(s)$ (Fig.~\ref{gFs_c}) is stronger for the sphere than for the half-torus.

\subsection{Field in $ab$-plane direction}
 
In order to calculate the characteristic function $g_F(s)$ for field directed
in $ab$-plane direction it is necessary to take into account the vortex
lattice distortion via the parameter $\tau$. For each value of the magnetic
field $\tau$ has to be found by minimizing the free energy of the system as 
described in section~\ref{spectrum}.
This makes the calculation of $g_F(s)$ more involved in this case. However,
even without knowing the value of $\tau$ one can get already insight into the
shape of the characteristic function $g_F(s)$ by analyzing the extrema and
saddle points of $|\eta_{k_F}|$ in Eqs. (\ref{etaabcyl}) and (\ref{etaabsph}).
Via Eq.~(\ref{gfs}) these saddle points will lead to van Hove singularities
in $g_F(s)$.

\begin{figure}
  \begin{center}
    \includegraphics[width=0.9\columnwidth]{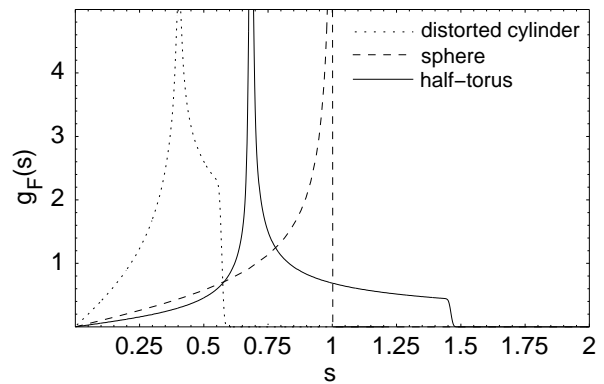}
    \vspace{.4cm}

    \caption{The characteristic functions $g_F(s)$ for the cylindrical, spherical 
    and half-toroidal Fermi surface. The magnetic field is applied in
    $ab$-plane direction. \label{gFs_ab} }
  \end{center}
\end{figure} 

For the cylindrical Fermi surface we find from Eq.~(\ref{etaabcyl}) that there
is a minimum at $|\eta_{k_F}|(\varphi=\pi/2,k_c=0)=0$, a maximum at
$|\eta_{k_F}|(\varphi=0,c k_c=\pi/2)/\alpha=\sqrt{e^{2\tau}+ e^{-2\tau} \epsilon_c^2 }$,
and two saddle points at $|\eta_{k_F}|(\varphi=0,k_c=0)/\alpha=e^{\tau}$ and
$|\eta_{k_F}|(\varphi=\pi/2,c k_c=\pi/2)/\alpha=e^{-\tau} \epsilon_c$ leading
to logarithmic singularities. 
In this case we can show analytically that $\tau$ is minimized by
$e^{\tau}=\sqrt{\epsilon_c}$ for all field strengths. Therefore, the
two saddle points collapse into one.
The characteristic functions obtained after minimization of $\tau$ are shown
in Fig.~\ref{gFs_ab} for field in $ab$-plane direction. The dotted line
shows the result for the distorted cylinder. The logarithmic van Hove
singularity due to the two saddle points is seen at
$s=\sqrt{\epsilon_c} \simeq 0.4$ and a step due to the maximum is found at
$s=\sqrt{2 \epsilon_c} \simeq 0.57$.

For the other three Fermi surfaces from Eq.~(\ref{etaabsph}) we find a
minimum at $|\eta_{k_F}|(\varphi=\pi/2,\vartheta=0)=0$ and two extremal
points at $|\eta_{k_F}|(\vartheta=\pi/2)/\alpha=e^{-\tau}$ and
$|\eta_{k_F}|(\varphi=0,\vartheta=0)/\alpha=e^{\tau}$. The point at 
$|\eta_{k_F}|/\alpha=e^{-\tau}$ for $\tau>0$ turns out to be an extended
saddle point, leading to a square-root singularity in $g_F(s)$.
In the case of a spherical or elliptical Fermi surface $\tau=0$ or
$\bar{\tau}=0$ is found due to symmetry, respectively. As we have noted
before, both the gap amplitude and the free energy functional 
depend on $\tau$ only via $|\eta_{k_F}|$. Also, as we have shown in
Eq.~(\ref{etaabsell}), we can bring 
$|\eta_{k_F}|$ of the elliptic Fermi surface into the same form as $|\eta_{k_F}|$
of the spherical Fermi surface with $\tau$ substituted by $\bar{\tau}$ and 
$v_F$ substituted by $\sqrt{v_{F,ab} v_{F,c}}$. In this case the two extremal 
points again collapse into one and we just rediscover the same function $g_F(s)$ as
for field in $c$-axis direction, as expected. For the half-torus the parameter
$\tau$ is nonzero and field dependent, however. Therefore, a square-root 
singularity is seen in
Fig.~\ref{gFs_ab} at $s=e^{-\tau}$ and a step due to the maximum at $s=e^{\tau}$.
Here, $g_F(s)$ is shown for high fields close to $B_{c2}$ where we find $\tau=0.375$.

\begin{figure}
  \begin{center}
    \includegraphics[width=0.9\columnwidth]{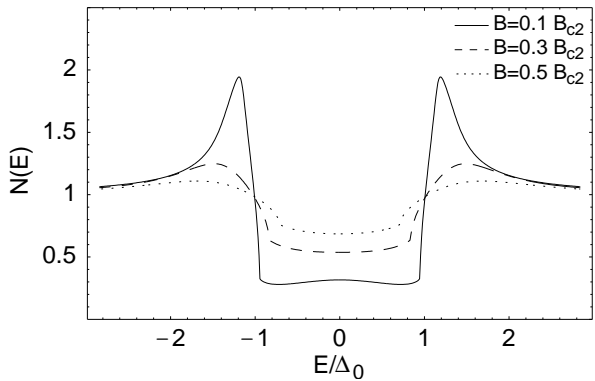}
    \vspace{.4cm}

    \caption{ Averaged density of states for the distorted cylinder with
    magnetic field in $ab$-plane direction 
    \label{ds_cyl_ab} }
  \end{center}
\end{figure} 

\begin{figure}
  \begin{center}
    \includegraphics[width=0.9\columnwidth]{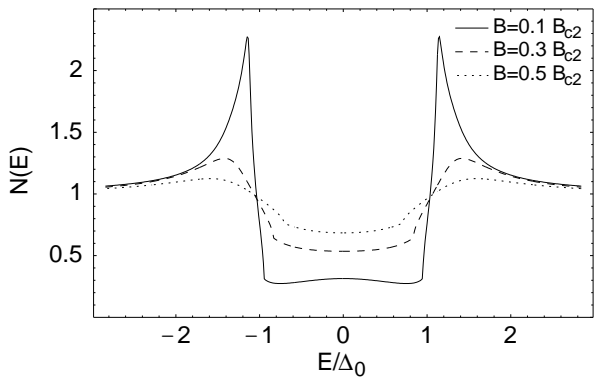}
    \vspace{.4cm}

    \caption{ Averaged density of states for the half-torus with
    magnetic field applied in $ab$-plane direction 
    \label{ds_torus_ab} }
  \end{center}
\end{figure} 

In Figs. \ref{ds_cyl_ab} and \ref{ds_torus_ab} we show the
corresponding densities of states for cylindrical and toroidal
Fermi surface, respectively. For this field direction the densities
of states for spherical (Fig.~\ref{ds_sphere}), distorted
cylindrical (Fig.~\ref{ds_cyl_ab}), and toroidal (Fig.~\ref{ds_torus_ab})
Fermi surfaces appear to be much more similar to each other. This
becomes clear from the similarity of the characteristic functions
in Fig.~\ref{gFs_ab}: all curves increase linearly at low values
of $s$. Again, the reason for this is topological: for field applied
in $ab$-plane direction in all three cases there exist a few singular
points on the Fermi surfaces at which the Fermi velocity is directed
parallel to the magnetic field, in contrast to the case of field
applied in $c$-axis direction. Accordingly, the results in 
Figs. \ref{ds_cyl_ab} and \ref{ds_torus_ab} resemble that of the 
spherical Fermi surface and we can identify again
the characteristic break 
at the gap edge and the distinct peaks on each side of the gap that we 
have already recognized in Fig. \ref{ds_sphere}.

\section{Comparison with a broadened BCS type density of states}
\label{comparison}

In analysis of experimental tunneling data on superconductors in the
vortex state a simple model of a BCS type density of states with a broadening
parameter $\Gamma$ of the following form has been used frequently
\cite{Gonnelli,Bugoslavsky,Naidyuk,Naidyuk2,Dynes}
\begin{equation} 
N(E) = \mbox{Re } \left[\frac{E + i \Gamma}{\sqrt{(E + i \Gamma)^2 - \Delta^2}} \right]
\label{broad}
\end{equation}
for $E>0$. Here, $\Delta$ and $\Gamma$ have been used as field dependent fitting 
parameters. In this section we want to make a critical comparison of this model 
with our results and point out the limitations of this model.

\begin{figure}
  \begin{center}
    \includegraphics[width=0.9\columnwidth]{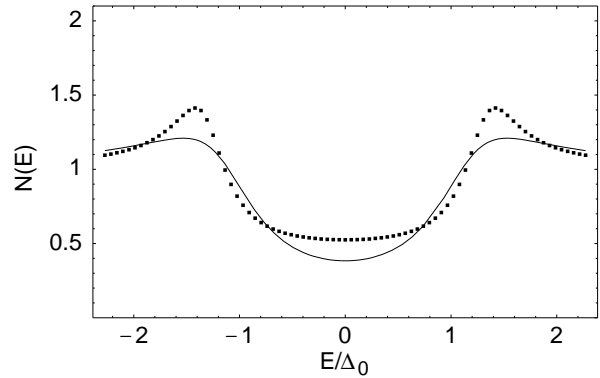}
    \vspace{.4cm}

    \caption{Spatially and momentum averaged density of states for $B=0.3 \: B_{c2}$.
	The boxes show the data from the quasiclassical calculation (also seen in 
        Fig.~\ref{ds_cyl}) and the solid line 
	represents the broadened BCS spectrum that was fitted to these data 
        using Eq.~(\ref{broad}).
     \label{fit_1} }
  \end{center}
\end{figure}

Looking at Figs.~\ref{ds_cyl}, ~\ref{ds_torus}, and ~\ref{ds_sphere} it is
immediately apparent that Eq.~(\ref{broad}) is certainly not able to reproduce
these very different shapes of the curves, because the results in
Figs.~\ref{ds_cyl}, ~\ref{ds_torus}, and ~\ref{ds_sphere} are shown for the
same reduced magnetic field and gap values (only the Fermi surface structure
was changed). Nevertheless one might ask whether these curves can be approximated
by Eq.~(\ref{broad}) with some effective broadening parameter $\Gamma$. In
order to answer this question we have made least squares fits of Eq.~(\ref{broad})
to our results for the cylindrical Fermi surface Eq.~(\ref{doscyl}) shown in 
Fig.~\ref{ds_cyl}. In Fig.~\ref{fit_1} we show the best fit for
$B=0.3 B_{c2}$. It appears that the width of the peaks at the gap edge
and the high value of the density of states at small energies cannot be
described simultaneously very well. The main reason for this failure is due
to the presence of a high amount of vortex core states that dominate at low
energies and cannot be captured very well by Eq.~(\ref{broad}).

\begin{figure}
  \begin{center}
    \includegraphics[width=0.9\columnwidth]{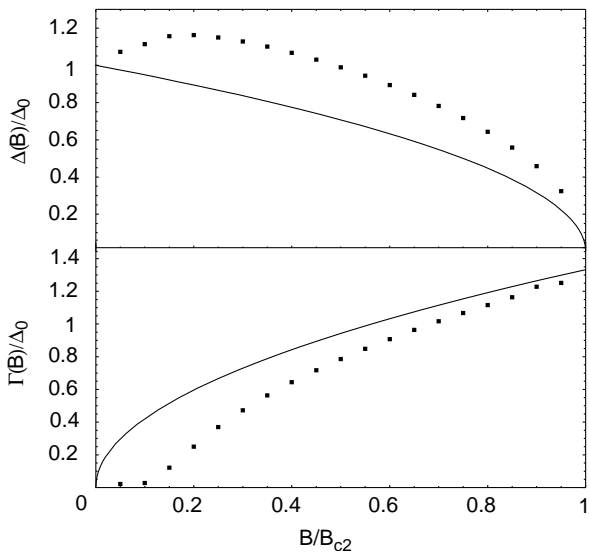}
    \vspace{.4cm}

    \caption{The upper panel shows the gap amplitude that was extracted 
	from the fitting process (boxes) in comparison with the field 
        dependence of the gap function
	that was used to calculate the quasiclassical spectra (solid line).
	The lower panel shows the variation of the fitting parameter $\Gamma$ (boxes) 
	as a function of the magnitude of the magnetic field compared with the 
        average Doppler shift $|\eta_{k_F}|/\sqrt{2}$ (solid lines). 
     \label{fit_2} }
  \end{center}
\end{figure} 

In Fig.~\ref{fit_2} we show the values of the two parameters
$\Delta$ and $\Gamma$ as a function of magnetic field found from our fits
(solid squares). In the upper panel we compare $\Delta(B)$ with the actual
field dependence of the gap shown as the solid line. This comparison shows
that the gap value extracted from Eq.~(\ref{broad}) does not very accurately
reproduce the actual gap value and care should be taken when interpreting
gap values extracted this way. The lower panel in Fig.~\ref{fit_2} shows the field
dependence of the broadening parameter $\Gamma$. It is clear that this quantity
increases with increasing magnetic field. For illustration we are also
showing the 'average Doppler shift' $|\eta_{k_F}|/\sqrt{2}$ as the
solid line. This quantity
represents the average energy broadening due to the local Doppler shift
of the energies of the excited quasiparticles in the presence of the
supercurrents around the vortices. The comparison shows that $\Gamma$
roughly measures this average Doppler shift, particularly at higher magnetic
fields.

Instead of using Eq.~(\ref{broad}) for an extraction of the magnetic field
dependence of the gap from experimental data it would be better to use
the theory presented here. This requires a priori knowledge of the
Fermi surface structure, however. For the special but important case of the 
cylindrical Fermi surface with field in $c$-axis direction the calculation 
simplifies and we can
suggest an improved version of Eq.~(\ref{broad}): first we note
that Eq.~(\ref{doscyl}) depends on
magnetic field only through the field dependence of the gap function
$\Delta(B)$ and the quantity $\alpha \propto \sqrt{B}$. Therefore,
Eq.~(\ref{doscyl}) can be brought into the following convenient
form:
\begin{equation}
N(B,E) = 
\mbox{Re} \left[  \frac{1}{\sqrt{1 + P_\Lambda (B, E)}} \right]
\label{dosapproxb}
\end{equation}
with
\begin{equation}
P_\Lambda(B, E) = 2 \frac{\bar{B}}{B} \left( \frac{\Delta(B)}{\Delta_0} \right)^2 
\left( 1+ i \sqrt{\pi} x w(x + i 0^+) \right)
\label{plambdab}
\end{equation} 
and
\begin{equation}
x= i z = \sqrt{ \frac{\bar{B}}{B}} \frac{E}{ \Delta_0}
\end{equation}
where we have introduced a characteristic magnetic field
\begin{equation}
\bar{B}=\frac{2 c}{\hbar v_F^2 e} \Delta_0^2
\end{equation}
such that $\sqrt{\frac{B}{\bar{B}}}= \frac{\alpha}{\sqrt{2}\Delta_0}$. 
In Eq.~(\ref{dosapproxb}) all
material dependent quantities including the Fermi surface structure
appear lumped into the single parameter $\bar{B}$, which makes 
Eq.~(\ref{dosapproxb}) particularly useful for fitting of experimental
data. Such a fitting could 
proceed as follows: first, the zero field value $\Delta_0$ has to be
extracted from a usual fit to zero field experimental data. Then,
Eq.~(\ref{dosapproxb}) can be fitted to finite field data using 
$\bar{B}$ as a field independent and $\Delta(B)/\Delta_0$ as a
field dependent fitting parameter. For this fitting procedure a 
useful approximation of the expression in Eq.~(\ref{plambdab}) 
containing Dawson's integral for real values of $x$ is given by
\begin{equation}
1+ i \sqrt{\pi} x w(x + i 0^+) \simeq \frac{1-x^2-0.2 x^4}{1+x^2+0.4x^6} + 
i \sqrt{\pi} x e^{-x^2}
\end{equation} 
This approximation is better than 2\% and reproduces the behavior
of the left hand side in the limit $x \rightarrow 0$ to order $x^2$ and 
in the limit $x \rightarrow \infty$ to order $1/x^2$.

In order to get some feeling for the parameter $\bar{B}$, we can relate
it to the upper critical field $B_{c2}$ using the linearized gap
equation. At $B_{c2}$ we find \cite{DahmSchopohl1}
\begin{equation}
\alpha(B_{c2})= \sqrt{2 \gamma} \Delta_0 \exp \left\{ - \left\langle 
\ln \frac{|\eta_{k_F}|}{\alpha}
\right\rangle_{FS} \right\}
\end{equation}
and therefore
\begin{equation}
\frac{\bar{B}}{B_{c2}} = \left( \frac{\sqrt{2}\Delta_0}{\alpha(B_{c2})} \right)^2 =
 \frac{\exp \left\{ 2 \left\langle 
\ln \frac{|\eta_{k_F}|}{\alpha}
\right\rangle_{FS} \right\} }{\gamma}
\label{bbar}
\end{equation}
For the cylindrical Fermi surface with field in $c$-axis direction considered
here $\frac{|\eta_{k_F}|}{\alpha}=1$ and therefore we have
$\bar{B}=B_{c2}/\gamma=0.561 B_{c2}$.

\section{\label{application} Application to MgB$_2$}

In this section we want to apply our findings to the situation in MgB$_2$.
It has been established recently that this compound belongs to the rare
case of a superconductor possessing two different sized gaps on two
different parts of the Fermi surface: a large gap living on the cylindrical
$\sigma$-bands and a small gap on the three-dimensional $\pi$-bands.
\cite{Liu,Louie,Jin,Eskildsen} The structure of the
$\pi$-band Fermi surface can be approximated by a half-torus as pointed out
in Ref. \onlinecite{DahmSchopohl1}. The quasiparticle excitations in the
vortex state of MgB$_2$ have been studied recently by tunneling spectroscopy
\cite{Gonnelli,Bugoslavsky,Ekino,Eskildsen} and the field dependence of the
superconducting gap was extracted. \cite{Gonnelli,Bugoslavsky,Ekino}

\begin{figure}
  \begin{center}
    \includegraphics[width=0.9\columnwidth]{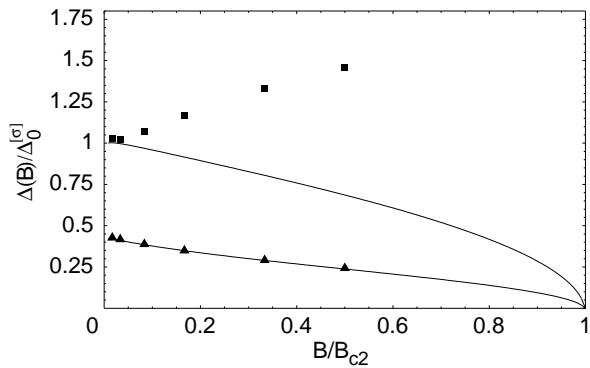}
    \vspace{.4cm}

    \caption{The two gaps of MgB$_2$ as a function of magnetic field. 
    The solid squares show the maxima of the peaks that result from the $\sigma$ band
    density of states, the triangles mark the flanks of the smaller gaps, 
    also extracted from the quasiparticle spectrum. 
     \label{twogaps} }
  \end{center}
\end{figure} 

The normalized total density of states of a two band superconductor is given
by a weighted average of the two partial densities of states for each of the
two bands:
\begin{equation}
N(B,E) = (1-w_\pi) N_{\sigma} (B,E) + w_\pi N_{\pi}(B,E)
\label{weightedsum}
\end{equation}
Here, $w_\pi$ is the weight of the $\pi$-band density of states and has been
calculated by band structure calculations \cite{Liu,Kortus} to be 
$w_\pi=0.577$. For the partial densities of states we can take
our results from section~\ref{integration} for the distorted cylinder with the 
distortion parameter $\epsilon_c=0.163$ and the half-torus $\nu=4$. 
The two different gap amplitudes have to be found from a solution of the
two by two gap equation in the vortex state:
\begin{equation}
\Delta^{\alpha}(\vec{r}) =  \pi T \sum_{\alpha'} \sum_{|\epsilon_n| < \omega_c} 
\lambda^{\alpha,\alpha'} \left\langle f^{\alpha'} (\vec{r},\vec{k}_{F_\alpha}, 
i \epsilon_n) \right\rangle_{FS}
\label{2by2_gap_eq}
\end{equation}
where we assumed an isotropic pairing interaction $\lambda^{\alpha,\alpha'}$
with $\alpha$ the band index: $\alpha = \sigma, \pi$.

\begin{figure}
  \begin{center}
    \includegraphics[width=1\columnwidth]{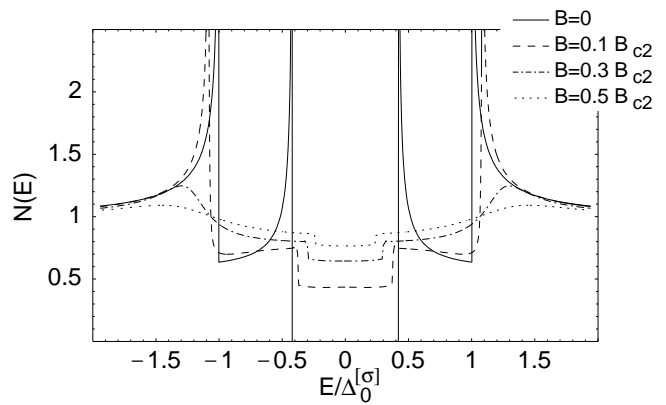}
    \vspace{.4cm}

    \caption{Averaged density of states for MgB$_2$ with magnetic field parallel to 
    the $c$-axis direction of the crystal lattice, calculated as a weighted sum of
    the averaged density of states of a toroidal Fermi surface and a cylindrical 
    Fermi surface using Eq.~(\ref{weightedsum}). Results are shown for
    different magnetic fields.
     \label{ds_MgB2} }
  \end{center}
\end{figure}

In Fig.~\ref{twogaps} the two gap amplitudes are shown as function
of the applied magnetic field for the set of parameters used in
Ref.~\onlinecite{DahmSchopohl1}. The resulting density of states for field
in $c$-axis direction is shown in Fig.~\ref{ds_MgB2}. From this figure
it becomes apparent that the peak at the gap edge of the small gap is
much more rapidly suppressed as the field is increased. This is due to
the topology of the $\pi$-band: as we have shown in the previous section
its half-torus shape leads to a much more rapid suppression of the peak
at the gap edge than for the cylindrical Fermi surface of the $\sigma$-band.
Experimentally, this effect has been noted by Gonnelli et al. \cite{Gonnelli}
Already at comparatively low fields of about 1 Tesla no apparent structure of the
$\pi$-band gap could be observed anymore, while the structure of the
$\sigma$-band gap remains visible to much higher fields of about 4-5 Tesla.

For comparison in  Fig.~\ref{twogaps} we are showing the position of the peak
in the $\sigma$-band density of states as a function of magnetic field as the
solid squares. The position of the sharp flanks in the $\pi$-band
density of states is shown as the triangles. Clearly, the structures
in the densities of states are behaving quite differently in the two bands
due to their different topologies. This effect should be taken into account
when analyzing the experimental data.

\section{\label{conclusion} Conclusions}

We studied the influence of Fermi surface topology on the density of
states in the vortex state of type II superconductors for the four Fermi
surface structures shown in Fig.~\ref{figfermisurf}. The topology of
the Fermi surface takes influence on the density of states, because the direction
of the Fermi velocity with respect to the local supercurrent flow around
the vortices leads to a change of the excitation energy of the quasiparticles.
We saw that the density of states behaves quite differently for cylindrical,
spherical and toroidal Fermi surfaces. The field dependence and shape of the curves
shows characteristic features related to the topology of the Fermi surface
in question. We showed that these features can be understood in terms
of characteristic Fermi surface functions. A particularly important role
is played by the number of points on the Fermi surface at which the
Fermi velocity is directed parallel to the external magnetic field.

We compared our results with the simple model of a broadened BCS type density of states,
that has been used frequently in the past for analysis of experimental data.
This comparison showed that the contribution coming from vortex core states
is underestimated in this simple model. This leads to inaccuracies in the
gap values extracted from the broadened BCS model. For the special case of
a cylindrical Fermi surface with field along the $c$-axis direction, we 
suggested an improved
formula for the density of states in the vortex state, in which all material
properties are lumped into a single parameter. This new formula does not
possess the limitations of the broadened BCS model and can be used
for fits to experimental data.

We applied our results to the case of the two gap superconductor MgB$_2$.
This case is particularly interesting, because the two Fermi surfaces related
to the two gaps possess completely different topology. We demonstrated that
this leads to very different field dependencies of the partial densities
of states, resembling recent observations on this compound.

\acknowledgments

We would like to thank M.~Eskildsen and R.~Gonnelli for valuable discussions.
S.~Graser acknowledges support through the 'Graduiertenf\"orderungsprogramm
des Landes Baden-W\"urttemberg'.

\appendix

\section{Calculation of the free energy}
\label{app1}

In this Appendix we provide a derivation of Eq.~(\ref{freeen}) for the free
energy difference within the method we are using here.
The difference of the free energy between the superconducting state
and the normal state for the case of an even parity superconductor can be calculated
by the coupling constant integration method \cite{AGD}
\begin{equation}
\Omega_S-\Omega_N = \int_0^1 \frac{d \lambda}{\lambda} \langle \lambda H_{int}\rangle_\lambda
\end{equation}
where $H_{int}$ denotes the interaction Hamiltonian. For a multiband superconductor this 
equation can be expressed
in terms of the gap function $\tilde{\Delta}^{\alpha} (\vec{r}, \vec{k}_F; \lambda)$ and the
anomalous Eilenberger propagator $\tilde{f}^{\alpha}(\vec{r}, \vec{k}_F, i \epsilon_n; \lambda)$
corresponding to a reduced pairing interaction 
$\lambda \cdot V^{\alpha \alpha^\prime}(\vec{k}_F,\vec{k}_F')$
\begin{eqnarray}
\Omega_S - \Omega_N = -\int_0^1 \frac{d \lambda}{\lambda} 
\int_{C_\Lambda} \frac{d^2 r}{|C_\Lambda|}
\sum_\alpha \left\langle \left[ \tilde{\Delta}^{\alpha} (\vec{r}, \vec{k}_F; \lambda) 
\right]^\dagger \right. \nonumber \\
\left. \times \pi T \sum_{|\epsilon_n| < \omega_c} 
\tilde{f}^{\alpha} (\vec{r}, \vec{k}_F, i \epsilon_n; \lambda) 
\right\rangle_{FS_\alpha}
\label{eqa2}
\end{eqnarray}
Here, $\alpha$ is the band index and $\tilde{\Delta}^{\alpha}(\vec{r}, \vec{k}_F; \lambda) $ is
solution of the renormalized gap equation
\begin{eqnarray}
\lefteqn{
\tilde{\Delta}^{\alpha} (\vec{r}, \vec{k}_F; \lambda) = } \label{gap_0} \\
& & \lambda \sum_{\alpha^\prime} 
\Bigg\langle V^{\alpha \alpha^\prime}  (\vec{k}_F,\vec{k}_F') 
\; \pi T \sum_{|\epsilon_n| < \omega_c} 
\tilde{f}^{\alpha^\prime} (\vec{r}, \vec{k}_F', i \epsilon_n; \lambda)
\Bigg\rangle_{FS'_{\alpha^\prime}}
\nonumber
\end{eqnarray}
Here, $\tilde{f}^{\alpha^\prime} (\vec{r}, \vec{k}_F', i \epsilon_n; \lambda)$ is a
solution of the Eilenberger equation Eq.~(\ref{eq1}) in the presence of the
$\lambda$-dependent gap function 
$\tilde{\Delta}^{\alpha} (\vec{r}, \vec{k}_F; \lambda)$ and depends on the 
parameter $\lambda$ only implicitly via $\tilde{\Delta}^{\alpha}$.
The integration over $\lambda$ in Eq.~(\ref{eqa2}) can be substituted by an
integration over the pairing potential, a usual technique as described in
textbooks.\cite{AGD} Generalizing this to the case of an inhomogeneous
superconductor we introduce a function $x(\lambda)$ via
\begin{equation}
x(\lambda) \cdot \Delta^{\alpha} (\vec{r}, \vec{k}_F) =
\tilde{\Delta}^{\alpha} (\vec{r}, \vec{k}_F; \lambda)
\label{xlam}
\end{equation}
Then, $\tilde{f}^{\alpha}$ can be viewed as a function of $x$, calculated for a reduced
gap $x \Delta^{\alpha}$, and the integration over $\lambda$ can be substituted
by an integration over $x$. In order to do so, we insert Eq.~(\ref{xlam}) into 
Eqs.~(\ref{eqa2}) and (\ref{gap_0}). Taking the derivative of Eq.~(\ref{gap_0}) with 
respect to $x$ we find
\begin{eqnarray}
\lefteqn{
\frac{x}{\lambda} \frac{d\lambda}{dx} \Delta^{\alpha} (\vec{r}, \vec{k}_F) = 
\Delta^{\alpha} (\vec{r}, \vec{k}_F) -} \label{dgap_0} \\
& & \lambda \sum_{\alpha^\prime} 
\Bigg\langle V^{\alpha \alpha^\prime}  (\vec{k}_F,\vec{k}_F') 
\; \pi T \sum_{|\epsilon_n| < \omega_c} 
\frac{d\tilde{f}^{\alpha^\prime}}{dx} (\vec{r}, \vec{k}_F', i \epsilon_n; x)
\Bigg\rangle_{FS'_{\alpha^\prime}}
\nonumber
\end{eqnarray}
Inserting the hermitian conjugate of Eq.~(\ref{dgap_0}) into Eq.~(\ref{eqa2})
and using the fact that the pairing interaction is hermitian, i.e.
\begin{equation}
\left[ V^{\alpha \alpha^\prime}  (\vec{k}_F,\vec{k}_F') \right]^\dagger =
V^{\alpha^\prime \alpha}  (\vec{k}_F',\vec{k}_F) 
\end{equation}
we arrive at
\begin{eqnarray}
\lefteqn{\Omega_S - \Omega_N = 
- \int_{C_\Lambda} \frac{d^2r}{|C_\Lambda|} \; \sum_\alpha} \\ & &
\Bigg\langle \pi T \sum_{|\epsilon_n| < \omega_c}
\int_0^1 dx \; \Bigg\{ \left[ \Delta^{\alpha} (\vec{r}, \vec{k}_F) \right]^\dagger 
\tilde{f}^{\alpha} (\vec{r}, \vec{k}_F, i \epsilon_n ; x)
\nonumber \\ & &
- \Delta^{\alpha} (\vec{r}, \vec{k}_F) 
\left[ x \frac{d\tilde{f}^{\alpha}}{dx} (\vec{r}, \vec{k}_F, i \epsilon_n ; x) 
 \right]^\dagger  \Bigg\} \Bigg\rangle_{FS_\alpha} 
\nonumber
\end{eqnarray}
A partial integration of the last term with respect to $x$ finally yields
\begin{eqnarray}
\Omega_S - \Omega_N = \int_{C_\Lambda} \frac{d^2r}{|C_\Lambda|} \; \sum_\alpha
\Bigg\langle \pi T \sum_{|\epsilon_n| < \omega_c}
\mbox{Re } \left\{ \left[ \Delta^{\alpha} (\vec{r}, \vec{k}_F) \right]^\dagger 
\right. \nonumber \\
\left. \left( f^{\alpha} (\vec{r}, \vec{k}_F, i \epsilon_n) 
 - 2 \int_0^1 dx \; \tilde{f}^{\alpha} (\vec{r}, \vec{k}_F, i \epsilon_n ; x) 
 \right) \right\} \Bigg\rangle_{FS_\alpha} 
\label{freeenergy}
\end{eqnarray}
This equation still holds generally for inhomogeneous multiband superconductors.

We can now employ the approximation procedure introduced in 
Ref.~\onlinecite{DGIS} and evaluate the spatial average in
Eq.~(\ref{freeenergy}). For a single band superconductor we find
\begin{eqnarray}
\frac{1}{|C_\Lambda|} \int_{C_\Lambda} d^2 r \; 
\left[ \Delta (\vec{r},\vec{k}_F) \right]^\dagger
\tilde{f} (\vec{r}, \vec{k}_F, i \epsilon_n; x) = \nonumber \\ 
\frac{|\Delta (\vec{k}_F)|^2}{\epsilon_n} \sqrt{\pi} z \, w (iz)
\frac{x}{\sqrt{1 + x^2 P_\Lambda (\vec{k}_F, i \epsilon_n)}}
\label{eqa6}
\end{eqnarray}
The first term in Eq.~(\ref{freeenergy}) is obtained for $x = 1$, of course. 
In this expression 
$z$ denotes the normalized Matsubara frequencies as introduced in Eq.~(\ref{zeps}). 
Using the same method Eq.~(\ref{gapequ}) can be derived from Eq.~(\ref{gap_0}) for
$\lambda = 1$ averaging over a unit cell of the vortex lattice and eliminating
the pairing interaction $V$ using the linearized gap equation at $T=T_c$.
Finally, the integration over $x$ in Eq.~(\ref{freeenergy}) can be carried out 
analytically using Eq.~(\ref{eqa6}) and we obtain the result in Eq.~(\ref{freeen}).

\end{document}